\documentclass[12pt]{article}
\usepackage[margin=25mm]{geometry}

\RequirePackage{amsthm,amsmath,amsfonts,amssymb}
\usepackage[sort]{natbib}
\RequirePackage[colorlinks,citecolor=blue,urlcolor=blue]{hyperref}
\RequirePackage{graphicx}
\usepackage{fancyhdr}
\usepackage{mathtools}
\usepackage{bbm}
\usepackage{epstopdf}
\usepackage{color}
\usepackage{listings}
\usepackage{booktabs,subfig,longtable}
\usepackage{array}
\usepackage{mathrsfs}
\newcolumntype{P}[1]{>{\centering\arraybackslash}p{#1}}

\usepackage{multirow}
\usepackage{comment}
\usepackage{xr}
\usepackage{caption} 
\usepackage[toc,page]{appendix}
\usepackage[utf8]{inputenc}
\usepackage[T1]{fontenc}
\usepackage{lmodern}
\usepackage{floatrow}
\usepackage[english]{babel}

\makeatother
\def\@fnsymbol#1{\ensuremath{\ifcase#1\or *\or \dagger\or \ddagger\or
   \mathsection\or \mathparagraph\or \|\or **\or \dagger\dagger
   \or \ddagger\ddagger \else\@ctrerr\fi}}
\newcommand{\ssymbol}[1]{^{\@fnsymbol{#1}}}
\newcommand{\RN}[1]{%
  \textup{\uppercase\expandafter{\romannumeral#1}}%
}

\newcommand*{\eop}{\hfill\ensuremath{\blacksquare}}%

\graphicspath{{plots/}}

\DeclareMathOperator*{\argmax}{arg\,max}

\DeclarePairedDelimiter\floor{\lfloor}{\rfloor}

\makeatletter
\newcommand*{\rom}[1]{\expandafter\@slowromancap\romannumeral #1@}
\makeatother

\makeatletter
\newcommand*{\addFileDependency}[1]{% argument=file name and extension
  \typeout{(#1)}
  \@addtofilelist{#1}
  \IfFileExists{#1}{}{\typeout{No file #1.}}
}
\makeatother

\newcommand*{\myexternaldocument}[1]{%
    \externaldocument{#1}%
    \addFileDependency{#1.tex}%
    \addFileDependency{#1.aux}%
}
\myexternaldocument{supp}

\newtheorem{theorem}{Theorem}[section]

\theoremstyle{remark}

\theoremstyle{plain}
\theoremstyle{definition}
\newtheorem{remark}{{\bf Remark}}
\newtheorem{assumption}{{\bf Assumption}}

\newcommand{\E}{\mathit{E}}

\newcommand{\trans}{^{\mbox{\tiny {\sf T}}}}

\def\R{\mathbb R}

\def\eop{\hfill $\Box$ \\ }

\def\nano{\scriptscriptstyle}
\newcommand\hi[1]{^{\nano #1}}
\def\real{\mathbb R}
\newcommand\ca[1]{{\cal{#1}}}
\newcommand\lo[1]{_{\nano #1}}

\def\loo#1{\lo {\mathrm{#1}}}

\def\cov{\mathrm{cov}}
\def\nano{\scriptscriptstyle}
\def\inv{\hi{\nano -1}}

\def\nano{\scriptscriptstyle}

\def\ka{\kappa}

\def\ali{&\,}

\def\mpinv{\hi{\dagger}}

\def\mpinv{\hi{\dagger}}

\def\N{\mathbb{N}}
\def\Pr{\mathbb{P}}
\def\mI{\mathbb{I}}

\def\convd{\stackrel{d}{\longrightarrow} }
\def\convP{\stackrel{\mathbb{P}}{\longrightarrow} }
\def\convw{\stackrel{\mathcal{L}}{\longrightarrow} }
\def\MMD{\mbox{MMD}}

\newcommand\inner[2]{\langle #1, #2 \rangle}

\newfont{\rsfsten}{rsfs10 scaled 1050}
\newfont{\rsfstena}{rsfs10 scaled 750}
\newfont{\rsfstenb}{rsfs10 scaled 800}

\usepackage{titlesec}

\titleformat{\section}
{\normalfont\large\bfseries}{\thesection}{1em}{}
\titleformat{\subsection}
{\normalfont\normalsize\bfseries}{\thesubsection}{1em}{}
\titleformat{\subsubsection}
{\normalfont\normalsize\bfseries}{\thesubsubsection}{1em}{}
\titleformat{\paragraph}[runin]
{\normalfont\normalsize\bfseries}{\theparagraph}{1em}{}
\titleformat{\subparagraph}[runin]
{\normalfont\normalsize\bfseries}{\thesubparagraph}{1em}{}

\newcommand\red[1]{{\color{red}{{#1}}}}

\def\E{{\rm E}\,}

\include{apalike}

\def\nat{\mathbb{N}}

\providecommand{\keywords}[1]
{
	\small	
	\textbf{\textit{Keywords---}} #1
}

\title{Kernel-based Method for Detecting Structural Break in  Distribution of Functional Data}
\author{Peijun Sang$^{1}$ and Bing Li$^{2}$ \\
	\small $^{1}$Department of Statistics and Actuarial Science, University of Waterloo \\
	\small $^{2}$Department of Statistics,  Penn State University \\
}
\begin{document}
\maketitle

\begin{abstract}
	We propose a novel method to detect and date  structural breaks in the entire distribution of functional data. Theoretical guarantees are developed for our procedure under fewer assumptions than in the existing work. In particular, we establish the asymptotic null distribution of the test statistic, which enables us to test the null hypothesis at a certain significance level. Additionally, the limiting distribution of the estimated structural break date is developed under two situations of the break size: fixed and shrinking towards 0 at a specified rate. We further propose a unified bootstrap procedure to construct a confidence interval for the true structural break date for these two situations. 
	These theoretical results are justified through comprehensive simulation studies in finite samples. We apply the proposed method to two real-world examples: Australian temperature data for detecting structural beaks and Canadian weather data for goodness of fit. 
\end{abstract}


% \newpage
% \spacingset{1.9} % DON'T change the spacing!
\section{Introduction}
In this article we consider structural break detection for functional data. In functional data analysis (FDA), there is a large literature on functional principal component analysis (FPCA) and functional regression models. Recent developments in these two main topics can be found in comprehensive monographs such as \cite{ramsay2005}, \cite{horvath2012} and \cite{hsing2015}. Structural break detection has been extensively studied in multivariate data. However, in the FDA, it has received less attention compared with the two aforementioned research topics. 

Early work on structural break detection for functional data mainly deals with random samples of independent functional data that can be written as a mean function plus 
identically and independently distributed errors. Without loss of generality, the random samples are assumed to be elements of $L^2[0, 1]$, and are generated as 
\begin{equation} \label{eq-DGP}
X \lo i (t) = \mu (t) + \mathbbm{1}(i \geq k \hi *) \delta (t) + \varepsilon \lo i (t),~~i = 1, \ldots, n.
\end{equation}
Here $k \hi * \in \{1, \ldots, n\}$ denotes the break date, $\delta(t)$ denotes the difference in the mean functions of $X \lo i$ before and after $k \hi *$, and $\varepsilon \lo i$'s are assumed to be i.i.d random errors with mean 0. 
\cite{berkes2009} developed a procedure for testing the null hypothesis of no structural breaks against the alternative of a single break in the mean function under model \eqref{eq-DGP}. They established the consistency of the estimator of $k \hi * / n$ under the alternative, but did not develop a method to quantify the uncertainty of the estimator. 
\cite{aue2009estimation} filled this gap by establishing the limiting distribution of the estimator, and \cite{aston2012detecting} and \cite{aston2012evaluating} relaxed the independence assumption of $\varepsilon \lo i$'s in model \eqref{eq-DGP}, and developed a similar procedure to detect and estimate $k \hi *$ in the presence of certain serial correlation in $\varepsilon \lo i$'s. 

It should be noted that the aforementioned procedures rely on dimension reduction techniques such as FPCA. 
On the one hand, the use of dimension reduction techniques converts this infinite-dimensional structural break detection problem to a finite-dimensional one. On the other hand, dimension reduction would result in a loss of information. In particular, as noted in \cite{aue2018}, when there exists a structural break and $\delta$ in model \eqref{eq-DGP} lies outside the space spanned by the retained functional principal components, a consistent FPCA-based test or break date estimator would be unavailable. To address this issue, 
%\red{\cite{sharipov2016sequential} proposed on   method based on the $L^2$ metric, and bootstrap procedure. and \cite{bucchia2017change} generalized this to a general Hilbert space.}
\cite{aue2018} developed a {\it fully functional} tool to detect the structural break based on the $L \hi 2$-norm of the functional cumulative sum (CUSUM) statistic. Moreover, they investigated the limiting distribution of the estimated break date and developed confidence intervals for $k \hi *$ based on the limiting distribution under the regime of a shrinking break size. \cite{li2024detection} proposed a procedure for detecting structural breaks in the mean function of high-dimensional functional time series, which are allowed to be cross-sectionally correlated over subjects and temporally dependent over time. 

Procedures for detecting structural breaks in the second moment of functional data have also been developed in the literature. In these procedures, functional data are still assumed to be generated from model \eqref{eq-DGP} with a structural break in the covariance function of $\varepsilon \lo i$'s but with $\delta(t) = 0$. 
For example, \cite{aue2020structural}
considered the structural break of the trace of the covariance operator. More general structural breaks of the covariance operator, e.g., structural breaks in the eigenvalues and/or eigenfunctions of the covariance operator, were studied in \cite{horvath2022change} 
and \cite{dette2021}. 
A summary of these methods for detecting the structural break in the mean or covariance function of functional data can be found in the monograph \cite{horvath2024change}.

However, structural breaks need not occur only through the means or the covariance operators of the functional data---indeed, it is easy to construct a series of random functions with no structural breaks in either their means or their covariance operators, but with a structural break in their distributions.
In this paper, we develop a novel procedure to detect the structural break in the underlying distributions of functional data, which has broader applications and is more powerful than the existing methods in detecting the structural break of functional data. This procedure is built upon the concept of {\it mean embedding} \citep{muandet2017kernel} and the structure of nested Hilbert spaces. Roughly speaking, we assume that the stochastic process, denoted as $X$, lies in a general Hilbert space, and then use a positive kernel $\kappa$ to construct a reproducing kernel Hilbert space (RKHS) defined on the first-level Hilbert space. Under mild conditions, the function $\E[\kappa(X, \cdot)]$, which is known as the mean embedding of $X$, can characterize the entire distribution of $X$. We make use of this fact to develop a procedure to detect the structural break in the distributions of a sequence of independently distributed random functions. In particular, we propose a specific 
functional CUSUM statistic to detect and estimate the structural break date in the spirit of fully functional approach in \citet{aue2018}. 

In addition to developing this procedure, we make three more important contributions. First, in contrast to the procedures mentioned above for detecting the structural break in the mean or covariance function, 
our framework does not require specific assumptions on the distribution of the sequence of stochastic processes as in \eqref{eq-DGP}. Second, we develop appealing theoretical properties for our procedure. More specifically, 
under mild conditions, we establish the limiting distribution of the proposed test statistic under the null hypothesis that there is no structural break. This allows us to detect the existence of a structural break in the distribution with theoretical guarantees. Moreover, in the presence of a structural break, we find the limiting distribution of the estimator of the structural break date under two regimes of the breaking size. Based on the limiting behavior, we construct a unified bootstrap confidence interval for the true structural break date. Third, we conduct extensive simulation studies to demonstrate the performance of our procedure compared with some alternative procedures. In particular, these studies show that the size and power of the proposed test statistic compare favorably with these alternatives under various simulation settings. Additionally, the true coverage probability of the bootstrap confidence interval is close to the nominal level under mild conditions. In contrast, the confidence interval developed in \cite{aue2018} is excessively conservative. 

Throughout the paper, $\convd$ denotes weak convergence of random variables taking values in $\R \hi p$ for some $p \in \nat$, while $\convw$ denotes weak convergence of random functions taking values in an infinite-dimensional metric space. For $i < j$, we use the notation $a\lo{i:j}$ to denote the vector $(a\lo i, \ldots, a\lo j)$. 
\begin{comment}
Given two arbitrary positive sequences $\{a \lo n: n \in \nat\}$ and $\{ b \lo n: n \in \nat \}$, we write $a \lo n \prec b \lo n$ if $a \lo n / b \lo n \to 0$, write $ a\lo n \succ b\lo n$ if $b \lo n \prec a \lo n$, write $a \lo n \preceq b \lo n$ if $a \lo n / b\lo n$ is a bounded sequence and write $a \lo n \asymp b \lo n$ if $a \lo n \preceq b \lo n$ and $b \lo n \preceq a \lo n$. 
\end{comment}
For a Hilbert-Schmidt operator $A$, $\|A\|\loo{HS}$ denotes its Hilbert-Schmidt norm. For any $x \in \real$, let $\floor*{x}$ be the largest integer that is smaller than or equal to $x$.

The remainder of the article is organized as follows. In Section \ref{sec-model}, we introduce relevant mathematical concepts for our method. In Section \ref{sec:method&theory} we develop the procedure to detect the structural break and estimate the break date. Theoretical properties of the proposed procedure are further investigated in Section \ref{sec:method&theory}. Section \ref{sec:implementation} provides additional details to implement our procedure. We carry out extensive simulation studies to demonstrate the performance of our procedure in Section \ref{sec:simulation}. In Section \ref{sec:real} we apply our procedure to two real-world examples. Section \ref{sec:conclusion} concludes the article. All theoretical proofs and some extra simulation results are relegated to the Supplementary Material.

\section{Problem Setup} \label{sec-model}

In this section, we introduce the structure of nested Hilbert spaces and the concept of mean embedding, which play important roles in our method development. 

% \subsection{Notation}

\subsection{Nested Hilbert spaces} \label{sec-nested}

Let $(\Omega, \ca F, P)$ be a probability space, $\mI$ a bounded and closed interval in $\real$, and $\ca H$ a separable  Hilbert space of functions on $\mI$. Without loss of generality, we assume that $\mI = [0, 1]$. 
Let $X: \Omega \to \ca H$ be a random element taking values  in $\ca H$  measurable with respect to $\ca F / \ca F \lo X$, where $\ca F \lo X$ denotes the Borel $\sigma$-algebra generated by the open sets in $\ca H$.  Let $P \lo X$ denote the distributions of $X$. 

Let $\ka: \ca H  \times \ca H  \to \real$ be a positive kernel and $\frak M \lo X$ be the RKHS generated by $\ka$.  We assume that $\ka$ is induced by the inner product in $\ca H$; that is, there exists a function $\rho : \R^3 \rightarrow \R^+$, such that for any $f, g \in \ca H$, 
$$
\ka(f, g) = \rho(\langle f, f \rangle\lo{\ca H},  \langle f, g \rangle\lo{\ca H},    \langle g,  g \rangle\lo {\ca H}).
$$
A typical example is the Gaussian radial basis (GRB) kernel 
\begin{equation} \label{eq-GRB}
\ka (f, g) = \exp (-\gamma \| f - g \|\lo{\ca H}\hi 2), 
\end{equation}
where $\gamma > 0$ is a tuning constant.  This is an extension of the GRB function with the Euclidean norm replaced by the $\ca H$-norm. Since the kernel of $\frak M \lo X$ is determined by the inner product of $\ca H$, we refer to $\frak M \lo X$ as the nested RKHS via $\rho$. Similar structures were adopted in \cite{li2017nonlinear} for nonlinear sufficient dimension for functional data and \cite{sang2022nonlinear} for nonlinear function-on-function regression.

\subsection{Maximum Mean Discrepancy} 
Let $\ca P$ denote the set of Borel probability measures defined on $\ca F \lo X$ and $\ca P \lo {\ka} \subset \ca P$ be the set of probability measures $P$ such that 
$\E\lo{X \sim P}[\ka(X, X)] < \infty$. We characterize the distance between two probability measures in $\ca P \lo \ka$ via mean embedding  \citep{muandet2017kernel}, which is the mapping
\begin{align*}
P \lo {\ka} \to \frak M \lo X, \quad    P \lo X \mapsto \int \lo {\ca H} \ka  (\cdot, x) d P \lo X (x). 
\end{align*}
We denote the right-hand side by $\mu  \lo X $. Note that
\begin{align*}
    \| \mu \lo X  \| \hi 2 \lo {\frak M \lo X} = \ali   \langle   \E \ka (X, \cdot), \E \ka (X, \cdot ) \rangle \lo {\frak M \lo X } \\
    = \ali  \E   \langle  \ka (X, \cdot), \E \ka (X, \cdot ) \rangle \lo {\frak M \lo X } \\
\le  \ali  \E   \langle  \ka (X, \cdot), \ka (X, \cdot) \rangle \lo {\frak M \lo X } \hi {1/2} \,  \langle \E \ka (X, \cdot ), \E \ka ( X, \cdot )  \rangle \lo {\frak M \lo X } \hi {1/2} \\    
=  \ali  \E  [   \ka (X, X) \hi {1/2} ] \,   \,  \| \mu   \lo X  \|  \lo {\frak M \lo X }.  
\end{align*}
Hence $ \| \mu   \lo X  \| \lo {\frak M \lo X} < \infty$, which means 
 $\mu  \lo X $ is a member of $\frak M \lo X$. 

Let $P\lo Y \in \ca P \lo \ka$ denote the distribution of another random element $Y$ in $\ca H$. Suppose that $\{X_1, \ldots, X_n\}$ and $\{Y_1, \ldots, Y_m\}$ are i.i.d samples from $P\lo X$ and $P\lo Y$, respectively.  
The {\it maximum mean discrepancy} (MMD) was introduced by \cite{gretton2012kernel} to test $P \lo X = P \lo Y$. In general, assume that $U$ and $V$ are random variables taking values in a same topological space $\ca S$  and $P$ and $Q$ denote their distribution, respectively. Let $\E\lo{U} \{f(u)\} :=\E\lo{U  \sim P} \{f(u)\}$ and $\E\lo{V}\{f(v)\} := \E \lo{ V \sim Q}\{f(v)\}$. For a class of functions $f: \ca S \rightarrow \R$, denoted by $\ca F$, the MMD is defined as
\begin{equation} \label{eq-MMD}
\MMD[\ca F, P, Q] = \sup\lo{f \in \ca F} [\E\lo{U} \{f(u)\} - \E\lo{V}\{f(v)\}]. 
\end{equation}
\cite{gretton2012kernel} suggested taking $\ca F$ as the unit ball of an RKHS. 

Based on \eqref{eq-MMD}, by Lemma 4 of \cite{gretton2012kernel}, we have
\begin{equation*}
	\MMD\hi 2 [\ca F, P\lo X, P\lo Y] = \|\mu\lo{X} - \mu\lo{Y}\|\lo{\frak M \lo X} \hi 2,
\end{equation*}
where $\ca F$ is the unit ball of $\frak M \lo X$. If the mean embedding under the reproducing kernel $\ka$ is injective, MMD is indeed a metric over $\ca P \lo {\ka}$.
According to \cite{sriperumbudur2011}, this amounts to saying $\ka$ is {\it characteristic} of $\ca P \lo {\ka}$; namely, two probability measures $P \lo X, P \lo Y \in \ca P \lo{\ka}$ satisfy that $P \lo X = P \lo Y$ if and only if $\mu \lo X = \mu \lo Y$.
By Theorem 3 of  \cite{wynne2020kernel}, the GRB kernel defined in \eqref{eq-GRB} is characteristic of $\ca P \lo {\ka}$.
Interested readers may refer to \cite{fukumizu2007kernel} and \cite{sriperumbudur2011} for more details about characteristic kernels.

Furthermore, it can be easily shown that $\MMD\hi 2$ can be rewritten as 
\begin{equation*} 
%\label{eq-sqMMD}
	\MMD\hi 2 [\ca F, P\lo X, P\lo Y] = \E\{\ka(X, X\hi ')\} - 2\E\{\ka(X, Y)\} + \E\{\ka(Y, Y\hi ')\},
\end{equation*}
where $X\hi '$ and $Y \hi '$ are identical and independent copies of $X$ and $Y$, respectively. According to \cite{wynne2020kernel}, when $n = m$, an empirical version of 
$\MMD\hi 2 [\ca F, P\lo X, P\lo Y]$ is given by a U-statistic:
\begin{equation} \label{eq:emMMD2}
\widehat{\MMD}\hi 2(X\lo{1:n}, Y\lo{1:n})  = \frac{1}{n(n - 1)} \sum\lo{i \neq j} \hi n h(Z\lo i, Z \lo j),
\end{equation}
where $Z\lo i = (X\lo i, Y\lo i)$ and $h(Z\lo i, Z\lo j) = \ka(X\lo i, X \lo j) + \ka(Y\lo i, Y \lo j)
- \ka(X\lo i, Y \lo j) - \ka (X\lo j, Y \lo i)$. 

We assume that $X_1, \ldots, X_n$ are  independently distributed random functions taking value in $\ca H$. 
The problem of primary interest in this article is to leverage $\MMD\hi 2$ to test the null hypothesis that there's no structural break in the distribution of this sequence. Moreover, if this null hypothesis is rejected, we aim to develop a method to date this structural break.

\subsection{Covariance operator}

\begin{comment}
\begin{assumption} \label{ass:vardef}
	There exists a constant $C > 0$ such that for any $f \in \frak M \lo X$, $\E[f^2(X_i)] \leq C \|f\|_{ \frak M \lo {X \lo i}}^2$, for any $1 \leq i \leq n$. 
%	\label{ass-var}
\end{assumption}

Let $L_2(P\lo {X})$ denote the class of all measurable functions of $X$ such that $\E[f^2(X)] < \infty$ under $P \lo X$. 
Assumption \ref{ass:vardef} ensures that the inclusion mapping $\frak M \lo X \rightarrow L_2(P\lo {X})$, $f \mapsto f$ is a bounded linear operator. 
Further, the bilinear form $\frak M \lo X \times \frak M \lo X \rightarrow \R$, $(f, g) \mapsto \cov(f(X), g(X) )$ is bounded and linear. Therefore under Assumption \ref{ass:vardef}, there must exist an operator 
$\Sigma \lo {X_iX_i}  \in \ca B (\frak M \lo {X \lo i})$, the collection of bounded and linear operators on $\frak M \lo {X\lo i}$, such that $\inner {f} {\Sigma \lo {X_iX_i}  g}\lo {\frak M \lo X } = \cov(f(X_i), g(X_i) )$. 
\end{comment}

\begin{assumption} \label{ass:iid}
	The $n$ random functions $X\lo 1, \ldots, X \lo n$ are independently and identically distributed with a random function $X$. 
\end{assumption}

\begin{assumption}\label{ass:finitemoment} \quad 
	  $\E[ \ka (X, X ) ] < \infty$. 
\end{assumption}

Obviously, Assumption \ref{ass:finitemoment} is met when $\ka$ is the GRB kernel. 
Let $L \hi 2(P\lo {X})$ denote the class of all measurable functions of $X$ such that $\E\{f\hi 2(X)\} < \infty$ under $P \lo X$. Assumptions \ref{ass:iid} and \ref{ass:finitemoment} ensure that the inclusion mapping $\frak M \lo X \rightarrow L \hi 2(P\lo {X})$, $f \mapsto f$ is a bounded linear operator. Indeed, since $f(X \lo i) = \inner{f}{\ka(X\lo i, \cdot)}\lo{\frak M \lo X}$ for any $f \in \frak M \lo X$, the Cauchy-Schwarz inequality leads to $\E\{f\hi 2(X \lo i)\} \leq \|f\|\lo { \frak M \lo X}\hi 2 \E[\ka(X\lo i, X \lo i)]$. 
Furthermore, by the same argument, the bilinear form $\frak M \lo X \times \frak M \lo X \rightarrow \R$, $(f, g) \mapsto \cov(f(X), g(X) )$ is bounded. Therefore under Assumptions \ref{ass:iid} and \ref{ass:finitemoment}, there exists a linear  operator 
$\Sigma \lo {XX}  \in \ca B (\frak M \lo {X})$, the collection of bounded and linear operators on $\frak M \lo {X}$, such that $\inner {f} {\Sigma \lo {XX}  g}\lo {\frak M \lo X } = \cov(f(X), g(X) )$. 
% Assumption \ref{ass:vardef} must hold under Assumption \ref{ass:finitemoment}. 

Under Assumptions \ref{ass:iid} and \ref{ass:finitemoment}, we can easily show that $\Sigma \lo {XX}$ is a trace-class operator. Then there exits a non-increasing and summable sequence of non-negative eigenvalues $\{\theta\lo \nu\}$  and corresponding eigenfunctions $\{\varphi\lo \nu\} \subset \frak M \lo X$  such that
$
\Sigma \lo {XX} = \sum\lo{\nu = 1} \hi {\infty} \theta\lo{\nu} (\varphi\lo \nu \otimes \varphi\lo \nu),  
$
where  $\inner{\varphi\lo \mu}{\varphi \lo \nu} \lo {\frak M \lo X} =  \delta\lo{\mu \nu}$, and $\delta \lo {\mu \nu}$ is the Kronecker $\delta$ function which equals  $1$ when $\mu = \nu$ and 0 otherwise. 
Further, $\ka(X, \cdot)$ admits the following representation:
\begin{equation} \label{eq-KLexpansion}
	\ka(X, \cdot) = \mu\lo X + \sum\lo{\nu = 1} \hi \infty \theta\lo{\nu} \hi{1/2} \xi\lo{\nu} \varphi\lo \nu(\cdot),
\end{equation}
where $\xi\lo\nu = \theta\lo{\nu} \hi{-1/2}\inner{\ka(X, \cdot) - \mu\lo X}{\varphi\lo \nu}\lo{\frak M \lo X}$, are mean zero and uncorrelated random variables.

\section{Methodology and Theoretical Results}
%\subsection{Fully observed trajectory} \label{sec:fulltrajectory}
\label{sec:method&theory}
In this section, we develop the procedure to detect the structural break in the distributions of $X\lo 1, \ldots, X \lo n$ based on the MMD defined in \eqref{eq-MMD}. If our proposed detection procedure suggests that there is a structural break, we further construct a confidence interval for the break date. Theoretical properties of the procedure are also established.

The basic idea of detecting a structural break is as follows. Suppose there is a structural break in the distribution of $X \lo i$'s at $k$. Namely, the distribution of $X\lo{1:k}$ is different from that of $X\lo{(k+1): n}$.  Based on the concept of MMD and the empirical version of MMD$^2$ in \eqref{eq:emMMD2}, we may take
$
\|\widehat{\mu}\lo{1:k} - \widehat{\mu}\lo{(k + 1):n} \| \lo {\frak M \lo X} \hi 2,
$
where $\widehat{\mu}\lo{1:k} = k\inv \sum\lo{i = 1} \hi k \ka(X\lo i, \cdot)$ and $\widehat{\mu}\lo{(k + 1):n} = (n-k)\inv \sum\lo{i = k + 1} \hi n \ka(X\lo i, \cdot)$ with a characteristic kernel $\ka$, as the test statistic. But as pointed out by \cite{berkes2009}, 
statistics based on this type of summations tend to be  inflated artificially when the break point $k$ is close to 1 or $n$, which corresponds to early break and late break. 
Therefore, we consider a CUSUM-type statistic instead: 
\begin{equation} \label{eq-fCUSUM}
S\lo k = \biggl\|\sum\lo{i = 1} \hi k \ka(X\lo i, \cdot) - \frac{k}{n}\sum\lo{i = 1} \hi n \ka(X\lo i, \cdot) \biggl\| \lo {\frak M \lo X} \hi 2.
\end{equation}
If there is indeed a structural break at $k$, the value of $S \lo k$ tends to be relatively large. 
%where $\widehat{\mu} = n \inv \sum\lo{i = 1} \hi n \ka(X\lo i, \cdot)$. 
Then for any $\tau \in (0, 1)$, setting $k = \floor*{n\tau}$, we obtain a distance metric
$$
d \lo n(\tau) = \biggl\|\sum\lo{i = 1} \hi {\floor*{n\tau}} \ka(X\lo i, \cdot) - \frac{\floor*{n\tau}}{n}\sum\lo{i = 1} \hi n \ka(X\lo i, \cdot) \biggl\| \lo {\frak M \lo X} \hi 2.
$$

\begin{remark} \label{re-testst}
Alternatively, instead of calculating $d\lo n (\tau)$ directly, we can project the difference of these two empirical means onto particular directions. In the context of detecting the structural break in the mean function of independently distributed functional data, \cite{berkes2009} chose the first $d$ FPCs of $X_i$'s as the projection directions. In our context, we can perform kernel FPCA on $X\lo i $'s to estimate $\varphi\lo \nu$'s defined in \eqref{eq-KLexpansion}. However, as pointed out by \cite{aue2018}, the use of a fully functional test statistic has potential advantages over the projection-based method in detecting and dating the mean break of $X_i$'s. Therefore, we adopt $d_n(\tau)$ for detecting the structural break in the distribution of $X \lo i$'s. 
\end{remark}

Let $\{B\lo{\nu}(\cdot): \nu \in \N\}$ be a sequence of independent standard Brownian bridges. The following theorem establishes weak convergence for the stochastic process $\{d\lo n (\tau): 0 \leq \tau \leq 1\}$ under the null hypothesis, where there is no structural break in the distribution of these independent random functions.

\begin{theorem}\label{thm:fullweakconvergence}
	Under Assumptions \ref{ass:iid} and \ref{ass:finitemoment}, 
	$$
	\frac{d\lo n(\tau)}{n} \convw \sum\lo{\nu = 1} \hi{\infty} \theta \lo {\nu} B \hi 2\lo{\nu}(\tau) \qquad \tau \in [0, 1], 
	$$
in the Skorohod topology of $D[0, 1]$, a collection of functions on $[0, 1]$ that are right-continuous and have left limits. 
\end{theorem}

By Theorem \ref{thm:fullweakconvergence}, we consider a Kolmogorov–Smirnov-type procedure for testing structural breaks \citep{hong2024kolmogorov}. In particular, define
\begin{equation} \label{eq-KS}
T_n = \sup_{\tau \in [0, 1]} \frac{d \lo n(\tau)}{n}. 
\end{equation}
By the continuous mapping theorem, one has, under Assumptions \ref{ass:iid} and \ref{ass:finitemoment}, as $n \rightarrow \infty$, 
$$
T_n  \convd  \sup \lo{\tau \in [0, 1]} \sum\lo{\nu = 1} \hi{\infty} \theta_{\nu}  B\lo{\nu} \hi 2 (\tau). 
$$
Then we reject the null hypothesis at significance level $\alpha$ if $T_n > C\lo{\alpha}$, where $C\lo{\alpha}$ satisfies $\Pr \left(T_n > C\lo{\alpha}\right) = \alpha$. Implementing this test requires estimation of $C\lo{\alpha}$, which will be discussed in Section \ref{sec:implementation}.

Next we investigate the asymptotic behavior of $T \lo n$ under the alternative hypothesis. To avoid obscuring the main idea by technical complexities due to multiple structure break dates, we will focus on the scenarios with a single structural break date. This motivates us to consider the following scenario under the alternative hypothesis.

\begin{assumption} \label{ass:change}
The $n$ random functions $X \lo 1, \ldots, X \lo n $ are independently distributed. 
For some $k\hi * = \floor*{n\tau\hi *}$ with $\tau \hi * \in (0, 1)$,
	the probability laws of $X\lo 1, \ldots, X\lo {k \hi *}$ are identical and denoted as $P\lo 1$, while the probability laws of $X\lo{(k\hi * + 1)}, \ldots, X_n$ are identical and denoted as $P \lo 2$, with $P_1 \neq P_2$. Further, $\E[\{\ka(X\lo i, \cdot) - \mu \lo 1\} \otimes \{\ka(X\lo i, \cdot) - \mu \lo 1 \}] = \E[\{\ka(X\lo j, \cdot) - {\mu \lo 2} \} \otimes \{\ka(X\lo j, \cdot) - {\mu \lo 2} \}]$   for any $i \in 1:k \hi *, j \in (k\hi * + 1):n$, where $\mu\lo 1$ and $\mu\lo 2$ denote the mean embeddings of $P \lo 1$ and $P\lo 2$, respectively. 
\end{assumption}

\begin{theorem} \label{thm:TnHa}
Under Assumptions \ref{ass:finitemoment} and \ref{ass:change}, 
$T \lo n \convP  \infty$ as $n \rightarrow \infty$. 
\end{theorem}

Theorem \ref{thm:TnHa} states that under the alternative hypothesis specified in Assumption \ref{ass:change}, $T \lo n$ would diverge in probability. 
If the null hypothesis is rejected, define the estimator of the structural break date as
$$
\hat{k}\lo n = \min\left\{k: S \lo k = \max\lo{1 \leq k  ' \leq n} S \lo {k  '} \right\}. 
$$
Under Assumption \ref{ass:change}, we still use $\Sigma\lo{XX}$ to denote the common covariance operator of $P\lo 1$ and $P \lo 2$. 
Let
$\epsilon \lo i = \kappa(X \lo i, \cdot) - \mu \lo 1$ for $1 \leq i \leq k \hi *$ and $\epsilon \lo i = \kappa(X \lo i, \cdot) - \mu \lo 2$ for $k \hi * < i \leq n$. 
% Similar to \eqref{eq:partsum}, 
Obviously, the mean embedding of $\epsilon \lo i$ for $1 \leq i \leq n$ is 0 and its covariance operator is $\Sigma\lo{XX}$ under Assumption \ref{ass:change}. 
Let $\delta = \mu \lo 2 - \mu \lo 1$ denote the break size. 
As in \cite{aue2018}, we study the asymptotic behavior of $\hat{k} \lo n - k \hi *$ under two situations: the fixed break size situation where $\delta$ is independent of $n$, and the shrinking break size situation where $\delta = \delta \lo n$ converges to 0 at a specified rate. Let $\hat{\tau} \lo n = \hat{k} \lo n/n$. The following theorem establishes the consistency for $\hat{\tau} \lo n$, and the limiting distribution for the structural break date estimator under Assumption \ref{ass:change} with a fixed break size.
\begin{theorem} \label{thm:dateestconsistency}
%If the difference in the two mean embeddings, $\mu\lo 1 - \mu\lo 2$, is not in the orthogonal complement of the space spanned by the first $p$ eigenfunctions of $\Sigma\lo{XX}$, 
Under Assumptions \ref{ass:finitemoment} and \ref{ass:change} with $\delta \neq 0$, we have as $n \rightarrow \infty$, 
\begin{itemize}
\item[(i)] $\hat{\tau} \lo n \convP \tau \hi *$;
\item[(ii)] \begin{equation} \label{eq:weakcon-kest}
\hat{k} \lo n - k \hi * \convd \min \left\{k: G(k) = \sup\lo{k' \in \mathbb{Z}}G(k ')\right\}, 
\end{equation}
where 
$$
G(k) = \begin{cases}
(1 - \tau \hi *) \|\delta\|\lo{\frak M \lo X} \hi 2 k + \inner{\delta}{V \lo k} \lo {\frak M \lo X}, & ~~k < 0,\\
0,& ~~k = 0, \\
-\tau \hi * \|\delta\|\lo{\frak M \lo X} \hi 2 k + \inner{\delta}{V \lo k} \lo {\frak M \lo X},& ~~k > 0, 
\end{cases}
$$
with $V \lo k = \sum \lo {i = 1} \hi{k} \epsilon \lo i$ for $k > 0$ and $\sum \lo {i = k} \hi {-1} \epsilon \lo i$ for $k < 0$, where  $\epsilon \lo {-i}$  denotes an independent copy of 
% (resp. $\varepsilon \lo {i}$)
$\epsilon \lo i$  for $i > 0$. 
\end{itemize}

\end{theorem} 

According to Theorem \ref{thm:dateestconsistency}, the limiting distribution of $\hat{k} \lo n - k \hi *$ depends on $\epsilon \lo i$'s and $\delta$ when the break size $\delta$ is fixed. This fact renders great difficulty in constructing confidence intervals for the break date $k \hi *$. Inspired by \cite{aue2018}, we investigate the limiting distribution of $\hat{k} \lo n - k \hi *$ in the second situation where $\delta = \delta \lo n$ depends on the sample size and shrinks towards 0 at a specified rate.
We anticipate a more tractable limiting distribution that can facilitate the construction of confidence intervals for $k \hi *$. 

\begin{theorem} \label{thm:dateestconsistency2}
%If the difference in the two mean embeddings, $\mu\lo 1 - \mu\lo 2$, is not in the orthogonal complement of the space spanned by the first $p$ eigenfunctions of $\Sigma\lo{XX}$, 
Suppose that Assumptions \ref{ass:finitemoment} and \ref{ass:change} hold with $0 \neq \delta = \delta \lo n$, where $\delta \lo n$ satisfies $\|\delta \lo n\| \lo {\frak M \lo X} \rightarrow 0$ but $n \|\delta \lo n\| \lo {\frak M \lo X} \hi 2 \rightarrow 0$. Then 
we have,  as $n \rightarrow \infty$, 
\begin{itemize}
\item[(i)] $\hat{\tau} \lo n \convP \tau \hi *$;
\item[(ii)] 
\begin{equation*} \label{eq:weakcon-kest2}
 \|\delta \lo n\| \lo {\frak M \lo X} \hi 2(\hat{k} \lo n - k \hi *) \convd \inf \left\{x: U(x) = \sup \lo {x ' \in \real} U(x')\right\}, 
\end{equation*}
where 
$$
U(x) = \begin{cases}
(1 - \tau \hi *)x + \sigma W(x), & ~~x < 0,\\
-\tau \hi * x + \sigma W(x),& ~~x \geq 0, 
\end{cases}
$$
with $(W(x): x \in \real)$ a two-sided Brownian motion, and 
$$
\sigma \hi 2 = \lim \lo { n \rightarrow \infty} \frac{\inner{\Sigma \lo {XX} \delta \lo n}{\delta \lo n} \lo {\frak M \lo X} }{ \|\delta \lo n\| \lo {\frak M \lo X} \hi 2}. 
$$
\end{itemize}

\end{theorem} 

\begin{remark} \label{re-datelim}
When replacing $\ka(X \lo i, \cdot)$ with $X \lo i$ in \eqref{eq-fCUSUM} to detect the structural break in the mean function of a sequence of random functions, a similar result to
Part (i) of Theorem \ref{thm:dateestconsistency} was established in \cite{berkes2009}, and a similar result to Part (ii) of Theorems \ref{thm:dateestconsistency} and \ref{thm:dateestconsistency2} was established in \cite{aue2009estimation} and \cite{aue2018}. 
It should be noted that these reference works impose a specific assumption on the data generation process as in model \eqref{eq-DGP}, where $\varepsilon \lo i$' are assumed to be i.i.d copies of the error process \citep{berkes2009, aue2009estimation} or dependent innovations satisfying a certain dependence structure specified in Assumption 1 of \cite{aue2018}. In contrast, Theorems \ref{thm:dateestconsistency} and \ref{thm:dateestconsistency2} developed for our procedure do not require this stringent assumption, which greatly enhances the applicability of our method. More importantly, the proposed method can detect not only the structural break in the mean or covariance function of $X \lo i$'s, but also the change in the underlying distribution of $X \lo i$'s. The advantage of our method over these alternatives will be further demonstrated through simulation studies in Section \ref{sec:simulation}. 

We can construct a confidence interval for $k \hi *$ in the situation of a shrinking break size based on part (ii) of Theorem \ref{thm:dateestconsistency2}. Let $\hat{\sigma}$ and $\hat{\delta} \lo n$ denote the estimates of the nuisance parameter $\sigma$ and break size $\delta \lo n$, respectively. For any $\alpha \in (0, 1)$, 
we construct an asymptotic $(1 - \alpha)$-sized confidence interval for $k \hi *$ as 
\begin{equation} \label{eq:CI}
\left(\hat{k} \lo n - \frac{q \lo {1 - \frac{\alpha}{2}}}{\|\hat{\delta} \lo n\| \lo {\frak M \lo X} \hi 2}, \hat{k} \lo n + \frac{q \lo {\frac{\alpha}{2}}}{\|\hat{\delta} \lo n\| \lo {\frak M \lo X} \hi 2}\right), 
\end{equation}
where $q \lo {\alpha}$ denotes the $q$th quantile of $\Lambda = \inf \{x: U(x) = \sup \lo {x ' \in \real} U(x ')\}$. More details for the construction of this confidence interval can be found in Section \ref{sec:implementation}.

\end{remark}

\section{Sample-Level Implementations} \label{sec:implementation}
\subsection{Choice of tuning parameters in the characteristic kernal}
We use the GRB kernel defined in \eqref{eq-GRB}  for our numerical implementations, which require the choice of the tuning parameter $\gamma$. We extend the widely adopted median heuristic \citep{garreau2017large} to the Gaussian kernel defined on $\ca H \times \ca H$. In particular, we choose $\gamma$ as
$$
\frac{1}{\gamma} = \text{median}\{\|X \lo i - X \lo j\| \lo {\ca H} \hi 2: i, j = 1, \ldots, n, i \neq j\}. 
$$
Theoretical properties for this particular choice of $\gamma$ were investigated in \cite{wynne2020kernel} for functional data taking values in $L \hi 2(\mI)$, the space of square-integrable functions defined on $\mI$. 
In fact, in the literature of FDA, $L \hi 2(\mI)$ is widely adopted as the ambient space within which functional data are considered; see \cite{wang2016functional} for a detailed overview. 
In our numerical implementations, we take $\ca H = L \hi 2(\mI)$, but the proposed framework can be applied to other choices of $\ca H$. 

\subsection{Computation of critical values of the test statistics under null hypothesis}
To test the null hypothesis against the alternative one by using $T \lo n$ defined in \eqref{eq-KS}, an accurate estimate of the critical value of $T_n$ under $H \lo 0$, $C \lo {\alpha}$, is needed. 
According to Theorem \ref{thm:fullweakconvergence}, estimating $C \lo {\alpha}$ amounts to finding the estimated eigenvalues of $\Sigma \lo {XX}$. Under $H \lo 0$, we estimate $\mu \lo X$ and $\Sigma \lo {XX}$ as $\hat{\mu} \lo X = 1/n \sum \lo {i = 1} \hi n \kappa(X \lo i, )$ and the pooled covariance
$$
\hat{\Sigma} \lo {XX} = \frac{1}{n} \sum \lo {i = 1} \hi {n} \{\kappa(X \lo i, ) - \hat{\mu}\lo X\} \otimes \{\kappa(X \lo i, ) - \hat{\mu}\lo X\}. 
$$
Moreover, as pointed out by \cite{wang2012kernel} and \cite{li2017nonlinear}, when performing eigen-decomposition on $\hat{\Sigma}\lo{XX}$,
one should restrict their attention to ${\frak M} \lo X \hi 0$, which is spanned by $\{\ka(X\lo i, \cdot) - \hat{\mu}\lo X, i = 1, \ldots, n\}$, rather than the one spanned by $\{\ka(X\lo i, \cdot), i = 1, \ldots, n\}$. To this end, we employ the coordinate representation method for operators introduced in \cite{li2017nonlinear} to conduct the eigen-decomposition on $\hat{\Sigma}\lo{XX}$. 

Specifically, let $K\lo X = (\ka(X\lo i, X\lo j))\lo{ij}$ denote the $n \times n$ Gram matrix. Let $I_n$ and $1_n$ denote the $n \times n$ identity matrix and the $n$-dimensional column vector consisting of  1's, respectively. 
Then, by Proposition 3 of \cite{li2017nonlinear}, the coordinate representation for $\hat\Sigma\lo{XX}$ with respect to ${\frak M} \lo X \hi 0$ is 
$G \lo X = n\inv QK\lo X Q$, where $Q = I\lo n - 1_n 1_n \trans /n$. Any $\varphi \in {\frak M} \lo X \hi 0$  can be written as $\varphi = [\varphi] \trans Qb\lo X$, where $b\lo X = \{\ka(X\lo 1, \cdot), \ldots, \ka(X\lo n, \cdot)\}\trans$ and $[\varphi]$ denotes the vector of the coordinates of $\varphi$ with respect to $Qb\lo X$. 
Now we want to look for $\varphi \in {\frak M} \lo X \hi 0$ to maximize, subject to $\langle \varphi, \varphi \rangle \lo {\frak {M} \lo X} = 1$, 
$$
\inner{\varphi}{\hat{\Sigma}\lo{XX}\varphi}\lo{\frak M \lo X} = 
n\inv[\varphi]\trans  K\lo X QK\lo X Q [\varphi] = n [\varphi] G \lo X \hi 2 [\varphi], 
$$
where the last equality holds as $Q[\varphi] = [\varphi]$ by Proposition 2 of \cite{li2017nonlinear} and $Q^2 = Q$. 

Since $\inner{\varphi}{\varphi} \lo {\frak M \lo X} =
[\varphi]\trans G \lo X [\varphi]$, 
% and $\inner{\varphi}{\varphi\lo{\nu}} \lo {\frak M \lo X} = [\varphi]\trans G \lo X [\varphi\lo{\nu}]$ for any $\nu \geq 1$. 
letting $z = G\lo X \hi{1/2}[\varphi]$, we have $\inner{\varphi}{\varphi} \lo {\frak M \lo X} =
[\varphi]\trans G \lo X [\varphi] = 1$ if and only  $z \trans z = 1$. 
We consider the Moore-Penrose inverse of $G \lo X$, denoted as $G \lo X \mpinv$, to obtain $[\varphi]$ from $z$: $[\varphi] = (G\lo X \mpinv) \hi{1/2} z$. This representation leads to the following standard eigen-decomposition problem: for $\nu = 1, \ldots, p$,
$$
z\lo{\nu} = \argmax\lo {z \in \real \hi n} z \trans G\lo X G\lo X \mpinv G \lo X z,
$$
subject to $z\trans z = 1$ and $z\trans z\lo{k} = 0$ for any $k < \nu$. Namely, $z_1, \ldots, z_p$ are the first $p$ eigenvectors of the matrix $G\lo X G \mpinv \lo X G \lo X$. Consequently, $\hat{\theta} \lo {\nu}$, the estimated $\nu$th largest eigenvalue of $\Sigma \lo {XX}$, is that of $G \lo X$, and 
the corresponding estimated eigenfunction is  given by $\hat{\varphi}\lo{\nu} = z\lo{\nu}\trans (G \lo X \mpinv) \hi{1/2}Qb \lo X$ for $\nu = 1, \ldots, p$. 

To select $p$, the number of eigenvalues, to approximate the limiting distribution of $T \lo n$ under the null, we use the fraction of variance explained. That is, we choose $p$ as
$$
\hat{p} = \min\left\{p: \frac{\sum \lo {\nu = 1} \hi p \hat{\theta} \lo {\nu}}{\sum \lo {\nu \geq 1} \hat{\theta} \lo {\nu}} \geq 0.9 \right\}. 
$$
This strategy is widely adopted in the literature of FDA: see \cite{zhu2014structured}, \cite{aue2018} and \cite{hormann2022estimating},  for example. Alternatively, we can adopt the leave-one-curve out cross-validation method proposed by \cite{yao2005a}. 
Let $\hat{\kappa} \hi {(-i)}(X \lo i, )$ denote the predicted value for $\kappa(X \lo i, )$ based on \eqref{eq-KLexpansion} when the $i$th curve is removed from the preceding eigen-analysis.  
Then we choose optimal $p$ by minimizing the leave-one-out sum of residual squares:
$$
\sum\lo{i = 1} \hi n \sum\lo{j \neq i}  [\ka(X\lo i, X \lo j) - \inner{\hat{\kappa} \hi {(-i)}(X \lo i, \cdot )}{\kappa(X \lo j, \cdot )} \lo {\frak M \lo X}] \hi 2.
$$

\begin{comment}
We adapt the method proposed in Section 6.6 of \cite{li2017nonlinear} to choose the tuning parameter $\epsilon\lo X$. Let $[E\lo{XX}\hi{(-1)}] = (K\lo X \hi{(-i)} + \epsilon\lo X I\lo{n - 1}) \inv 
K\lo X \hi{(-i)}$, where $K\lo X \hi{(-i)}$ is submatrix of $K\lo X$ with the $i$th subject removed. 
Then we choose optimal $\epsilon\lo X$ to minimize the leave-one-out sum of residual squares:
$$
\sum\lo{i = 1} \hi n \sum\lo{j \neq i}  [\ka(X\lo i, X \lo j) - \{E \lo{XX}\hi{(-i)}\ka(X\lo j, \cdot)\}(X\lo i)] \hi 2.
$$
\end{comment}

\subsection{Confidence interval for the structural break date under alternative hypothesis} \label{subsec:bootCI}

In Remark \ref{re-datelim}, we develop an asymptotically $(1 - \alpha)$-sized confidence interval \eqref{eq:CI} for $k \hi *$ under Assumption \ref{ass:change} with a shrinking break size $\delta \lo n$. 
This confidence interval requires  accurate estimates of the nuisance parameters $\sigma$ and $\delta \lo n$. 

When the null hypothesis is rejected, given the estimated structural break date $\hat{k} \lo n$, we take the sample mean as the estimate of the two mean embeddings: 
$$
\hat{\mu} \lo 1 = \frac{1}{\hat{k} \lo n} \sum \lo {i = 1} \hi {\hat{k} \lo n} \kappa(X \lo i, )~~\text{and}~~\hat{\mu} \lo 2 = \frac{1}{n - \hat{k} \lo n} \sum \lo {\hat{k} \lo n + 1} \hi {n} \kappa(X \lo i, ). 
$$
Then $\hat{\delta} \lo n = \hat{\mu} \lo 2 - \hat{\mu} \lo 1$ is a natural estimate of $\delta \lo n$. Under Assumption \ref{ass:change}, we consider another pooled sample covariance for estimating $\Sigma \lo {XX}$ (with a slight abuse of notation):
$$
\hat{\Sigma} \lo {XX} = \frac{1}{n} \left[\sum \lo {i = 1} \hi {\hat{k} \lo n} \{\kappa(X \lo i, ) - \hat{\mu}\lo 1\} \otimes \{\kappa(X \lo i, ) - \hat{\mu}\lo 1\} + \sum \lo  {\hat{k} \lo n + 1} \hi n \{\kappa(X \lo i, ) - \hat{\mu}\lo 2\} \otimes \{\kappa(X \lo i, ) - \hat{\mu}\lo 2\}\right]. 
$$
Then an estimate of $\sigma \hi 2$ is immediately available. With $\hat{\sigma}$ and $\hat{\tau} \lo n$, we can easily estimate $q \lo {\alpha}$ by generating a sufficiently large number of i.i.d sample paths of the process $\{U(x): x \in \real\}$. 

Numerical studies demonstrate that the confidence interval developed above is excessively wide. Similar issues were identified in the confidence interval for the structural break date in the mean function in \cite{aue2018}. A plausible reason is that this confidence interval is built upon the assumption that the break size $\delta \lo n$ shrinks towards 0 at a specified rate, which is seriously violated in numerical studies. 
To address this issue, we adopt the bootstrap-based method proposed by \cite{antoch1995change}, who was the first to recognize the difference in the limiting behavior of $\hat{k} \lo n - k \hi *$ for independently distributed random variables with a mean change under these two situations: fixed $\delta$ and shrinking $\delta \lo n$. \cite{antoch1995change} proposed a unified bootstrap method to construct confidence intervals for $k \hi *$ for these two situations. In particular,
after we obtain the estimated structural break date $\hat{k} \lo n$, define the estimated residuals:
%\begin{singlespace}
$$
\hat{\epsilon} \lo i = 
\begin{cases}
\kappa(X \lo i, \cdot) - \hat{\mu} \lo 1,~~i = 1, \ldots, \hat{k} \lo n, \\
\kappa(X \lo i, \cdot) - \hat{\mu} \lo 2,~~i = \hat{k} \lo n + 1, \ldots, n.
\end{cases}
$$
%\end{singlespace}
\noindent Then we take $\epsilon \lo {1} \hi *, \ldots, \epsilon \lo n \hi *$ independently from the empirical distribution of $\hat{\epsilon} \lo 1, \ldots, \hat{\epsilon} \lo n$, and consider the bootstrap samples.
%\begin{singlespace}
$$
{Y} \lo i \hi * = 
\begin{cases}
\epsilon \lo {i} \hi * + \hat{\mu} \lo 1,~~i = 1, \ldots, \hat{k} \lo n, \\
\epsilon \lo {i} \hi * + \hat{\mu} \lo 2,~~i = \hat{k} \lo n + 1, \ldots, n.
\end{cases}
$$
%\end{singlespace}
\noindent Then we carry out the procedure in Section \ref{sec:method&theory} to estimate the structural break date by replacing $\kappa(X \lo i, \cdot)$ with $Y \lo i \hi *$ for $i = 1, \ldots, n$. We repeat this entire procedure  $B$  times,  and  employ the sample quanitles of these $B$ estimates of $k \hi *$ to construct confidence intervals for $k \hi *$. 

\section{Simulation Studies} \label{sec:simulation}

In this section, we perform extensive simulation studies to demonstrate the finite-ample performance of our proposed method in terms of detecting and dating structural breaks in the distribution of functional data. We compare our method with two alternative methods: the fully functional method proposed by \cite{aue2018} for detecting and dating structural breaks in the mean function (FF-mean for short), and the fully functional method 
proposed by \cite{horvath2022change} for detecting and dating structural breaks in the covariance function (FF-cov for short). While the two alternative methods focus on detecting the changes in mean and covariance function of the functional data, our method targets the entire distribution of the functional data. Our method is referred to as FF-dist for short. 

\subsection{Simulation designs}
Following the data-generation procedure in \cite{aue2018}, we generate functional data as
\begin{equation*}
X \lo i(t) = m \lo i (t) + \sum \lo {j = 1} \hi {21} \sqrt{\lambda \lo {ij}}\zeta \lo {ij} \phi \lo j (t),~~i = 1, \ldots, n, 
\end{equation*}
where $\phi\lo 1, \ldots, \phi\lo {21}$ denote the 21 Fourier basis functions on [0, 1] and $\zeta \lo {i1}, \ldots, \zeta \lo {i,21}$ denote the independent FPC scores. 
Each $X \lo i$ is observed at 50 equally-spaced time points on [0, 1]. 
Let $k \hi {*} = \floor*{n \tau \hi {*}}$ denote the true structure break date. We consider the following three settings. 
\begin{itemize}
\item[(a)] Setting 1: $\lambda \lo {ij} = j \hi {-1}$ for $j = 1, \ldots, 21$ and $i = 1, \ldots, n$; $m \lo i (t) = 0$  for $i \leq k\hi{*}$, while $m \lo i(t) = a \cdot \sum\lo{j = 1} \hi {21} \phi \lo j (t)/\sqrt{21} $ for $i > k \hi *$. As recommended by \cite{aue2018}, different values of $a$ are taken such that the signal-to-noise ratio, which is defined as
$$
\text{SNR} = \frac{a\tau \hi * (1 - \tau \hi *)} {\sum \lo {j = 1} \hi {21} \lambda \lo j},
$$
varies among 0, 0.1, 0.2, 0.3, 0.5, 1 and 1.5. Again, $\zeta \lo {ij}$'s are independently generated from $N(0, 1)$. 
\item[(b)] Setting 2: $m \lo i (t) = 0$ for $i = 1, \ldots, n$; $\lambda \lo {ij} = j \hi {-1}$ for $j = 1, \ldots, 21$ and $i \leq k \hi *$ while $\lambda \lo {ij} = c \cdot j \hi {-1}$ for $j = 1, \ldots, 21$ and $i >  k \hi *$. We take $c = 1, 1.6, 1.8, 2, 2.5 $ and 3 to accommodate slight to notable changes in the covariance operator after the structural break date. Moreover, $\zeta \lo {ij}$'s are independently generated from $N(0, 1)$. 
\item[(c)] Setting 3: $m \lo i (t) = 0$ and $\lambda \lo {ij} = j \hi {-1}$ for $i = 1, \ldots, n$ and $j = 1, \ldots, 21$. Moreover, $\zeta \lo {ij}$'s are independently generated from $N(0, 1)$ for $i \leq k \hi *$, while for $i > k \hi *$, $c \lo {\pi} \zeta \lo {ij}$'s are independently generated from a mixture model of the standard normal distribution and the student $t$ distribution with 3 degrees of freedom, i.e.,  $c\lo{\pi} \zeta \lo {ij} \sim \pi N(0, 1) + (1 - \pi)t \lo 3$, where $\pi$ denotes the mixing proportion and $c \lo {\pi}$ is chosen such that $var(\zeta \lo{ij}) = 1$. We take $\pi = 0, 0.2, 0.4, 0.6, 0.8$ and 1 to accommodate slight to notable changes in the distribution of $X$ after the structural break time. 
%\red{(Peijun, the notation $c \lo \pi$ didn't appear in the model for $X \lo i(t) $ introduced at the beginning of section 5.1).} \red{(Peijun: actually, we know the value of $c_{\pi}$: $c_{\pi}^2 = \pi + 3\pi$. This choice can guarantee that $\var(\zeta_{ij}) = 1$ in this mixture-model design for $\zeta_{ij}$'s.)}
\end{itemize}

\begin{figure}[H]
	\centering
   \includegraphics[width=\textwidth]{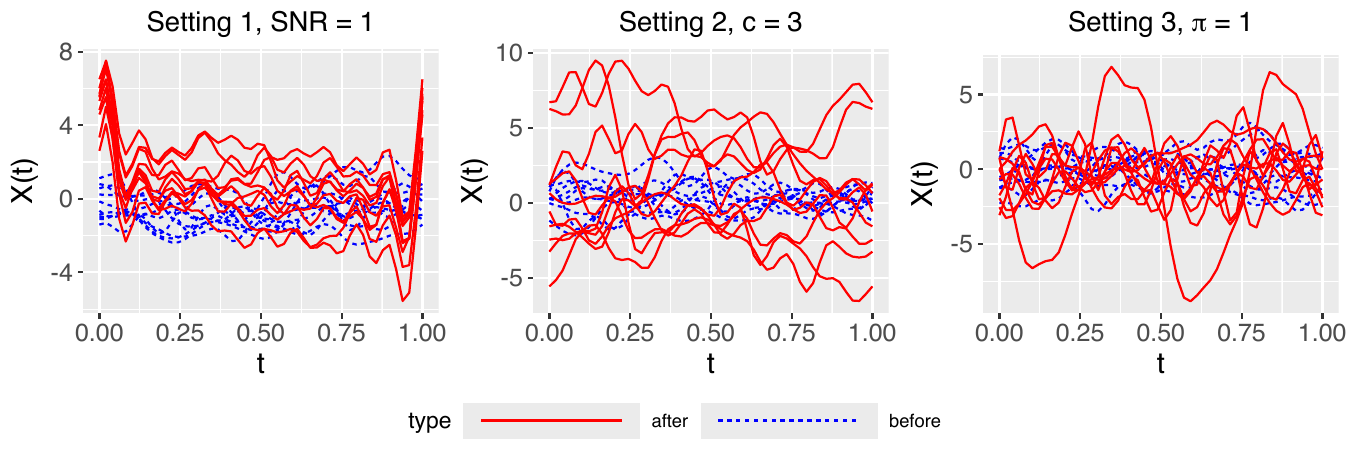}
   \caption{Profiles of functional data generated from the three settings. In each panel, red solid and blue dotted lines represent the trajectories of 10 randomly selected subjects after and before the structural break, respectively.}
	\label{fig:profiles}
\end{figure}

Setting 1 is exactly the same as setting 3 in \cite{aue2018}. Settings 1 and 2 are designed to showcase the performance of our method in detecting and dating the break time when there exists a structural break in the mean function and the covariance function of independently distributed functional data, respectively. In contrast to the previous two settings, there is no change in the mean function or covariance function of $X \lo i$'s in Setting 3. To accommodate a structural break in the underlying distribution of $X \lo i$'s, we consider a distributional change in the functional principal component (FPC) scores of $X \lo i$'s. In particular, we generate FPC scores from a mixture model as in \cite{yao2006penalized}. Figures \ref{fig:profiles} display the profiles of 10 randomly selected subjects before and after structural breaks for the three settings. We can easily identify a change in the mean function and the covariance function in the first two panels of Figure \ref{fig:profiles}. However, the difference between these groups of functional data is not discernible even with $\pi = 1$ in Setting 3 from Figure \ref{fig:profiles}. These findings indicate the escalated difficulty in detecting and dating the structural break in Setting 3 compared with the first two settings. 

To evaluate the performance of our method, we consider $\tau \hi * = 0$ (null hypothesis), and $\tau \hi * = 0.25 $ and 0.5, and sample size $n = 100$ and 200 in the first two settings, and $n = 500$ and 800 in Setting 3 because of the escalated difficulty. 1000 independent simulation trials are carried out for each simulation scenario to calculate relevant metrics. 

\subsection{Simulation results}
Table \ref{tab:size} summarizes the type-I errors of the three methods under the three simulations settings under various sample sizes. As expected, the empirical sizes of our method and its competitors are close to the nominal level for under all settings. These findings indicate that our method is not inferior to its competitors in terms of controlling type I errors.

\begin{table}[H] 
	\tabcolsep 0.3in
	\centering
	\caption{Empirical sizes of the detection methods under the three simulation settings with the nominal level $\alpha = 0.05$. }
{
	\begin{tabular}{@{}|ccccc|@{}}
	%	\toprule
	\hline 
		% &  
		Setting & $n$ & \multicolumn{3}{c|}{Methods}  \\
		\cline{1-5}
		& & FF-mean & FF-cov& FF-dist \\
	%	\midrule
		\multirow{2}{*}{1} & 100 & 0.084 &  & 0.058  \\
		
		& 200 & 0.076 &  & 0.070
		\\
		\hline
		\multirow{2}{*}{2} & 100 &  & 0.060 & 0.049  \\
		
		& 200 & & 0.058   & 0.051
		\\	
	  \hline

      \multirow{2}{*}{3} & 500 & 0.039  & 0.030 & 0.049  \\
		
		& 800 & 0.041 & 0.032   & 0.050
		\\	
	  \hline
	  \end{tabular}}
 % \vspace*{-5pt}
   	\label{tab:size}
\end{table}

\begin{figure}[H]
	\centering
   \includegraphics[width=\textwidth]{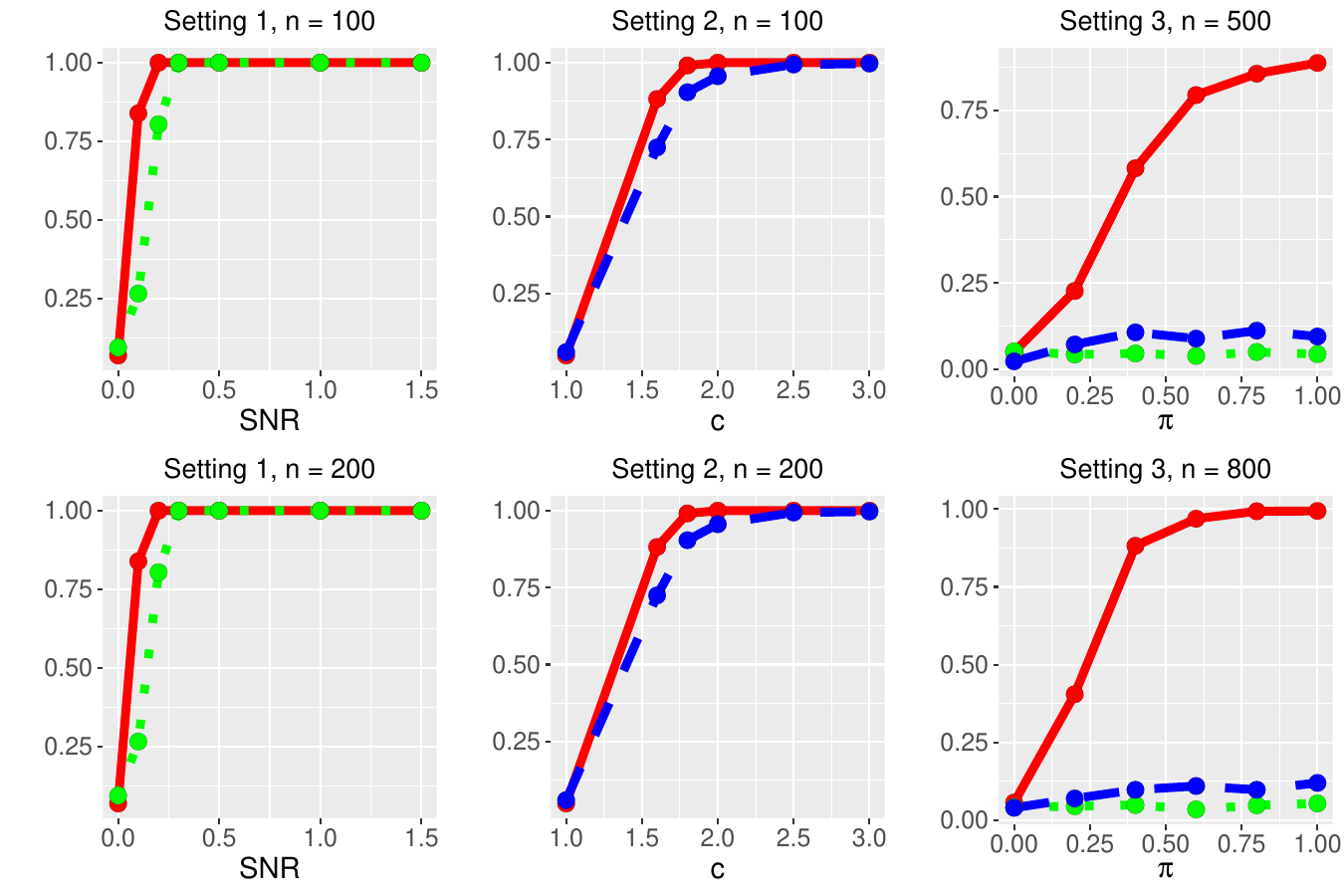}
   \caption{Power curves of the three detection methods under the three simulation settings with $\tau \hi * = 0.50$ for various sample sizes. In these panels, the red solid, green dotted and blue dashed lines represent the power curves of FF-dist, FF-mean and FF-cov, respectively.}
	\label{fig:power50}
\end{figure}

Figure \ref{fig:power50} displays the power curves of the three detection methods under the three simulation settings for different sample sizes with $\tau \hi * = 0.5$. Somewhat surprisingly, though FF-mean and FF-cov were designed to detect the structural break in the mean function and the covariance function, respectively, our method shows a greater power of rejecting the null hypothesis in the first two settings. The advantage of our method is more evident when the break size in the mean function or the covariance function is moderate. All these findings demonstrate the effectiveness of our method in detecting the structural break in the first two moments of independently distributed functional data. More noticeable advantages of our method over the two competitors are found in the two right panels of Figure \ref{fig:power50}, which showcase the power curves of these three methods in Setting 3. Since there is no structural break in either the mean function or the covariance function in Setting 3, FF-mean and FF-cov cannot reject the null hypothesis. However, our proposed method can effectively detect the structural break in the the underlying distribution of functional data, especially when the mixing proportion of the $t \lo 3$ distribution of the FPC scores becomes sufficiently large. To investigate whether these results are affected by the location of the structural break, we also consider $\tau \hi * = 0.25$ and the corresponding power curves are shown in Figure \ref{fig:power25} in the Supplementary Material. Similar patterns in the comparison of the power curves for these detection methods are identified in Figure \ref{fig:power25}.

\begin{comment}
In Setting 3, the empirical sizes of FF-mean and FF-cov are slightly smaller than the nominal level. This indicates that these two testing methods are  conservative when there is no change in either the mean function or the covariance function of functional data. This conclusion will be further justified in the power analysis as displayed in Figure \ref{fig:power50}. 
In contrast, our method still controls the type I error reasonably well when there is 
\end{comment}

\begin{figure}[H]
	\centering
   \includegraphics[width=\textwidth]{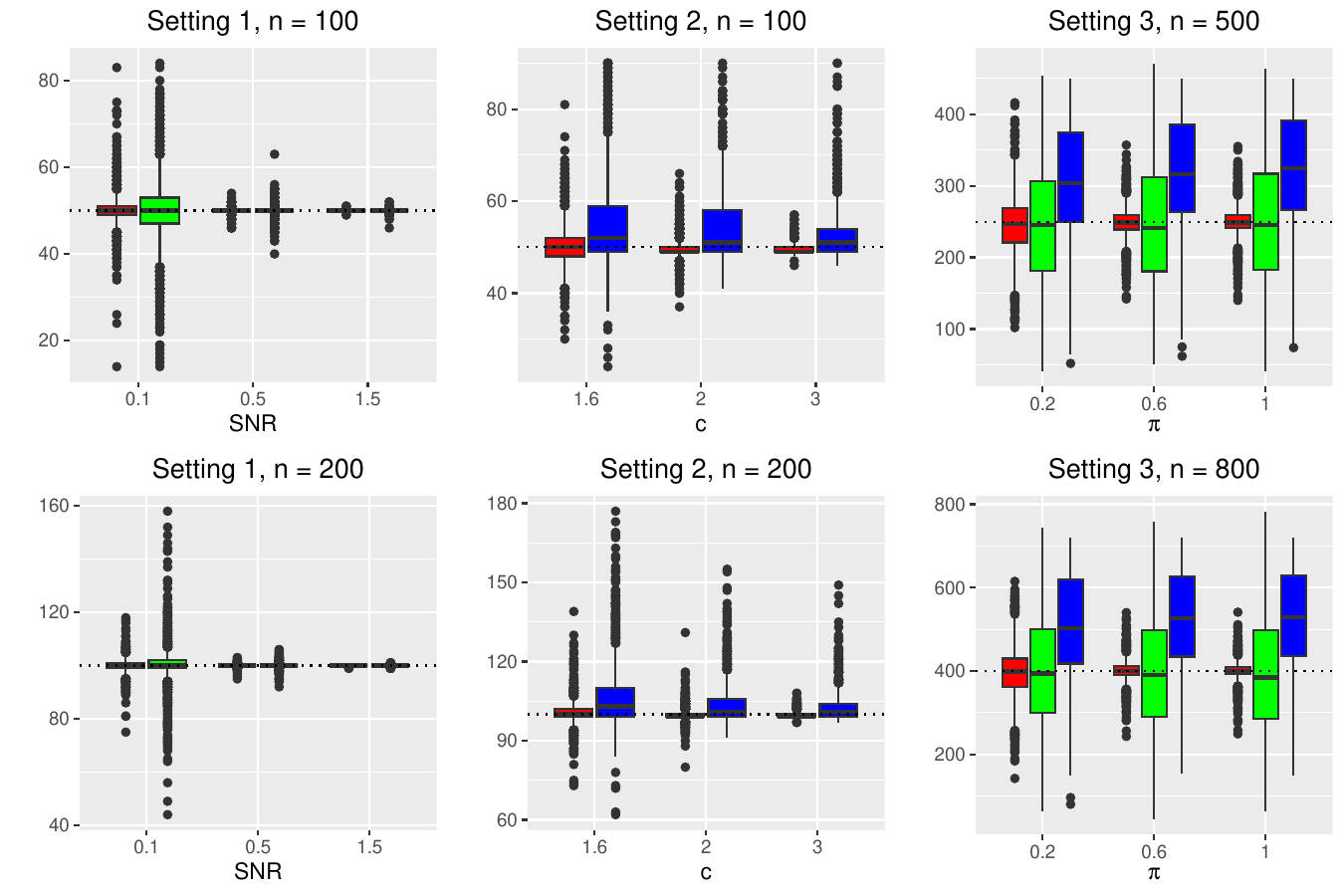}
   \caption{Boxplots of the estimated structural breaks for the three detection methods under the three simulation settings with $\tau \hi * = 0.50$ across 1000 Monte Carlo simulations. In these panels, the red, green and blue boxplots are for FF-dist, FF-mean and FF-cov, respectively. The dotted horizontal
 line represents the true structural break time.}
	\label{fig:box50}
\end{figure}

Figure \ref{fig:box50} depicts the boxplots of the estimated structural break dates obtained from these three methods across 1000 Monte Carlo simulations under these three settings for $\tau \hi * = 0.5$. In Settings 1 and 2, FF-dist and its competitors can date the true structural break time reasonably well. But our proposed method is more stable regardless of sample size or the break size of the mean function or the covariance function. Given that FF-mean and FF-cov were proposed to focus on detecting the structural break in the mean function and the covariance function, respectively, our method provides a unified and powerful framework to detect and date the structural breaks in these population parameters of functional data. In Setting 3, which seems to be the most challenging scenario for dating the structural break as indicated in Figure \ref{fig:profiles}, our method still shows remarkable performance in dating the structural break for relatively larger $\pi$ and $n$. If we take a closer look at such boxplots for $\tau \hi * = 0.25$ in Figure \ref{fig:box50} in the Supplementary Material, we find that, in contrast, FF-mean and FF-cov, especially the latter, cannot accurately date the structural break in this scenario. 

\begin{table}[!ht]
	\tabcolsep 0.2in
	\centering
	\caption{The empirical coverage probabilities of the 95\% bootstrap confidence intervals for $\tau \hi *$ under the three simulation settings with $\tau \hi * = 0.50$ based on 1000 Monte Carlo simulations.}
	\begin{tabular}[t]{c c c c c c c c}
		%\hline
		\hline
		Setting & $n$ & & & & & & \\
		\hline
		\multirow{4}{*}{1}& & &\multicolumn{5}{c}{SNR} \\[0.8ex]
		\cline{4-8}
		& & & $0.1$ & $0.2$ & $0.3$ & $0.5$ & $1$ \\[0.5ex]
		%\hline
		\cline{4-8}
		% &  & & & & & \\
		% &  & & & & & \\
		& $100$ & & 0.998 & 0.976 & 0.974 & 0.958 & 0.974 \\
		& $200$ & & 0.984 & 0.978 & 0.975 & 0.966 & 0.983 \\
		\hline
		\multirow{4}{*}{2}& & &\multicolumn{5}{c}{$c$} \\[0.8ex]
		\cline{4-8}
		& & & 1.6 & 1.8 & 2 & 2.5 & 3 \\[0.5ex]
		%\hline
		\cline{4-8}
		% &  & & & & & \\
		% &  & & & & & \\
		& $100$ & & 0.974 & 0.961 & 0.957 & 0.947 & 0.948 \\
		& $200$ & & 0.949 & 0.955 & 0.939 & 0.948 & 0.961 \\
		\hline
		\multirow{4}{*}{3}& & &\multicolumn{5}{c}{$\pi$} \\[0.8ex]
		\cline{4-8}
		& & & 0.2 & 0.4 & 0.6 & 0.8 & 1 \\[0.5ex]
		%\hline
		\cline{4-8}
		% &  & & & & & \\
		% &  & & & & & \\
		& $500$ & & 0.982 & 0.963 & 0.942 & 0.931 & 0.940 \\
		& $800$ & & 0.966 & 0.929 & 0.919 & 0.928 & 0.921 \\
		\hline
	\end{tabular}
	\label{tab:CP_tau50}
\end{table}

\begin{figure}[H]
	\centering
   \includegraphics[width=\textwidth]{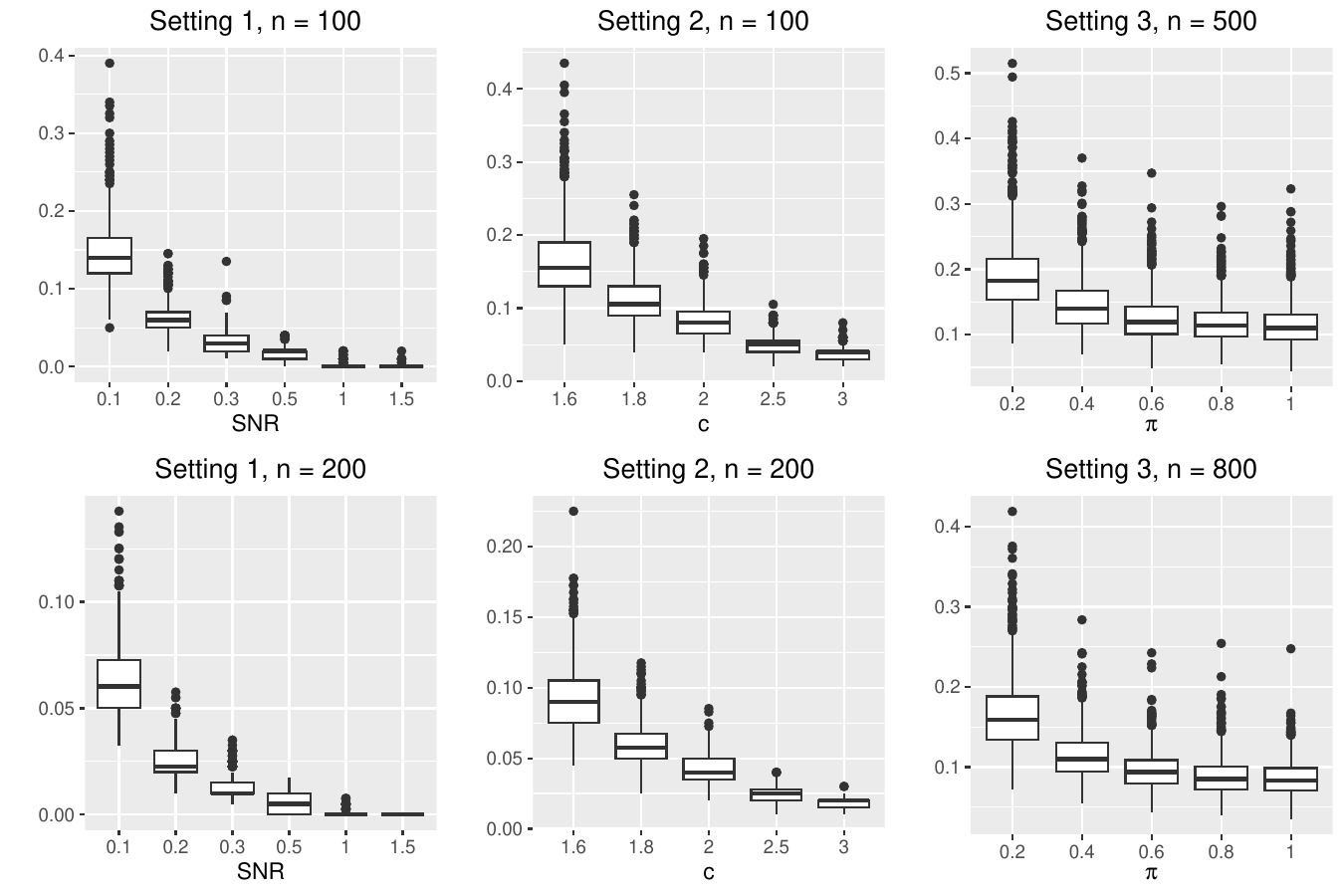}
   \caption{Boxplots of the lengths of the 95\% bootstrap confidence intervals for $\tau \hi *$ under the three simulation settings with $\tau \hi * = 0.50$ across 1000 Monte Carlo simulations.}
	\label{fig:CI_box50}
\end{figure}

Lastly, we examine the performance of the bootstrap confidence intervals developed in Section \ref{subsec:bootCI}. 
Figure \ref{fig:CI_box50} presents the boxplots of the lengths of the 95\% confidence intervals of $\tau \hi *$ for the three settings across 1000 Monte Carlo simulations, where $\tau \hi * = 0.5$ and we generate $B = 500$ bootstrap resamples in each simulation run. In Figure \ref{fig:CI_box50} we find that the confidence interval becomes shorter when the structural break size becomes larger or more samples are available in these three settings. Similar patterns are found in Figure \ref{fig:CI_box25} in the Supplementary Material, which presents the boxplots of the lengths of those confidence intervals with $\tau \hi * = 0.25$. We further assess the true coverage probability of the bootstrap confidence interval for $\tau \hi *$. Table \ref{tab:CP_tau50} summarizes the empirical coverage probabilities of these 95\% bootstrap confidence intervals for $\tau \hi * = 0.5$ based on 1000 simulation runs for the three simulation settings. When the sample size or the structural break size is reasonably large, the empirical coverage probability is close to the nominal level.

\section{Real Data Analysis} \label{sec:real}
In this section, we consider two data examples. In the first example,  we apply  our proposed method and  the two alternative methods  to  detect and date the structural break in the profiles of the temperature data collected in Australia. In the second example, we apply our procedure to the Canadian weather data  to evaluate the goodness-of-fit of linear function-on-function regression. 

\subsection{Australian temperature data}
\begin{figure}[H]
	\centering
   \includegraphics[width=\textwidth]{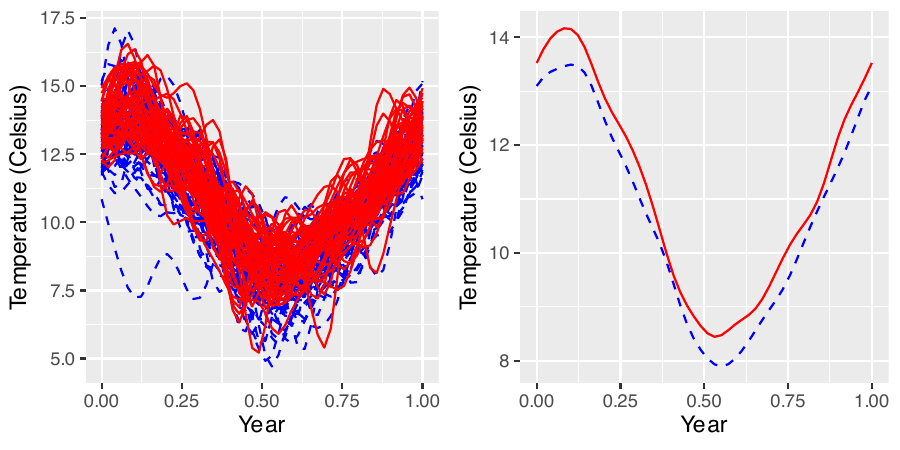}
   \caption{Left: profiles of minimum temperatures after and before the estimated structural break year based on FF-mean. Right: the mean functions of the temperature profiles after and before the estimated structural break. In both panels, the red solid and blue dashed lines represent the profiles from after and before the structural break, respectively.}
	\label{fig:temp_profiles}
\end{figure}

We first study the temperature data collected at eight weather stations in Australia. 
At each site, the raw data consist of 365 daily observations of minimum temperatures over a certain time span, which may vary from site to site. 
The sites and the corresponding time spans during which the data were collected are shown in the first two columns of Table \ref{fig:temp_profiles}. The raw data are available from the Australian Bureau of Meteorology at \url{www.bom.gov.au.}
Both \cite{aue2018} and \cite{chen2023greedy} have considered structural breaks for this dataset from the perspective of the FDA. 

\begin{table}[!ht]
\centering
\caption{Summary of the estimated structural breaks of the temperature curves and the corresponding 95\% confidence intervals based on FF-mean and FF-dist at these eight sites.}
\begin{tabular}[t]{c c c c }
%\hline
\hline
Station & Time Span & $\hat{k} \lo {\text{FF-mean}}$  & $\hat{k} \lo {\text{FF-dist}}$   \\
\hline
Boulia Airport & 1888-2012 &  1978 (1960, 1981) & 1978 (1967, 1979) \\
Cape Otway Lighthouse & 1864-2012 & 1970 (1920, 1976)  & 1970 (1946, 1972) \\
Gayndah Post Office & 1893-2009 & 1962 (1956, 1964) & 1962 (1959, 1963) \\
Gunnedah Pool & 1876-2011 & 1965 (1929, 1974) & 1965 (1958, 1966) \\
Hobart (Ellerslie Road) & 1882-2012 & 1965 (1959, 1967) & 1967 (1963, 1968) \\
Melbourne (Regional Office) & 1855-2012 & 1958 (1955, 1959) & 1958 (1956, 1958) \\
Robe Comparison & 1884-2012 & 1960 (1943, 1967) & 1972 (1957, 1974) \\
Sydney (Observatory Hill) & 1859-2012 & 1957 (1952, 1959) & 1957 (1954, 1958) \\
\hline
\end{tabular}
\label{tab:temperature}
\end{table}

As in \cite{aue2018} and \cite{chen2023greedy}, the raw observations collected in each year at each station are smoothed with 21 Fourier basis functions. Meanwhile, we remove years at which the proportion of missing values exceeds 1/3, from our study.
We only compare the performance of our procedure with that of FF-mean developed in \cite{aue2018}, but do not consider the procedure proposed in \cite{chen2023greedy}, since the latter focuses on detecting multiple structural breaks based on segmentation. Both our detection procedure and FF-mean yield tiny 
p-values at each of these 8 stations. These findings suggest that there exists a structural break in the mean function of the temperature profiles. 

We further employ FF-mean and FF-dist to estimate the structural break in the mean function or the distribution of the temperature curves at each of these eight stations. Table \ref{tab:temperature} summarizes the estimated structural breaks and the associated 95\% confidence intervals obtained from these two procedures. Except the stations Hobart and Robe Comparison, these two procedures give rise to identical estimated structural breaks, while our procedure always yields short confidence intervals, which echo those from the simulation studies. 

\begin{figure}[H]
	\centering
   \includegraphics[width=\textwidth]{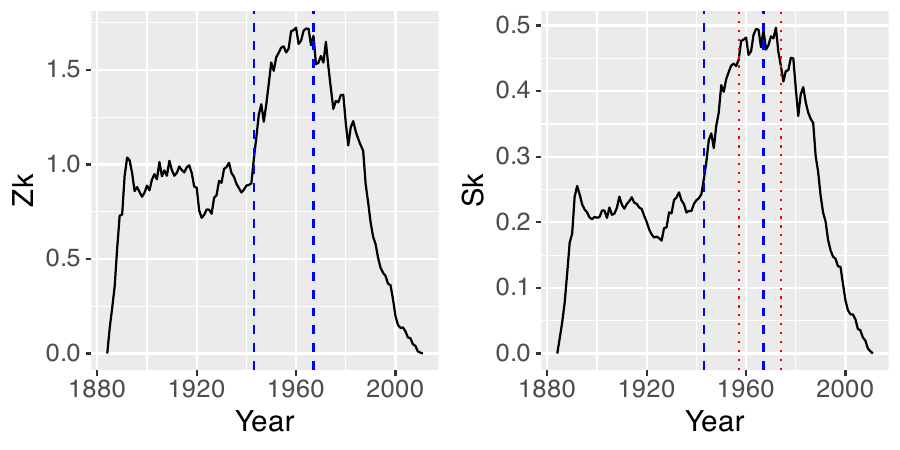}
   \caption{Trace of functional CUSUM statistics from FF-mean (left) and FF-dist (right).}
	\label{fig:temp_stat}
\end{figure}

Now we take a closer look at the estimated structure breaks obtained from these two procedures at Robe Comparison. Figure \ref{fig:temp_profiles} presents the profiles and mean functions of minimum temperatures after and before 1960, which is the structural break year obtained from FF-dist. The mean function of minimum temperatures after 1960 is slightly larger than that before 1960. To evaluate the brak size in the mean function compared with that in the distribution, we further calculate the functional CUSUM statistics of these two procedures, as shown in Figure \ref{fig:temp_stat}. It turns out that the confidence interval obtained from our procedure, consists of years that have relatively large functional CUSUM statistics defined in \eqref{eq-fCUSUM}. 
In contrast, the confidence interval from FF-mean contains years that have relatively small $Z_k$'s, the functional CUSUM statistics defined in \cite{aue2018}. Lastly, it should be noted that our confidence interval contains 1960, but the confidence interval from FF-mean does not contain 1972, which is the estimated structural break year from our procedure. These demonstrate the effectiveness and robustness of our procedure. 

%use the ACF or partial ACF in \cite{yeh2023functional} to test serial correlations. 

\subsection{Canadian weather data}
In the second data example we apply our proposed procedure in conjunction with a linear function-on-functionn regression method on the Canadian weather data, to showcase our method as a  goodness-of-fit test for function-on-function regression.  The data consist of 365 daily observations of temperature and precipitation at 35 different locations in Canada from 1960 to 1994. The dataset is available from the R package {\it fda} \citep{fda}. Moreover, this dataset has been widely studied in linear function-on-function regression, where both the response and the covariate are random functions; see Chapter 16 of \cite{ramsay2005} and \cite{sun2018optimal} for example. 

Let $Y \lo i(t)$ and $X \lo i(s)$ denote the profiles of daily precipitation and daily temperature in year $i$, respectively. We consider the following linear function-on-function regression model:
\begin{equation} \label{eq:ffreg}
Y \lo i(t) = \alpha(t) + \int \lo 0 \hi {365} X \lo i(s) \beta(t, s) ds + e \lo i(t),
\end{equation}
where $\alpha(t)$ and $\beta(t, s)$ represent the intercept and the slope functions, respectively, and $e \lo i(t)$'s are assumed to be i.i.d error processes.  
Efforts on the theoretical properties of fitting this model were devoted to establishing the convergence rate of an estimated slope function \citep{yao2005b} or mean prediction \citep{sun2018optimal}. In contrast, the goodness-of-fit for this model has received less attention; see \cite{gonzalez2022review} for an overview of research in this regard. In this application, we are interested in testing the null $e \lo i(t)$'s are i.i.d based on the $\hat{e} \lo i (t)$'s, which are residual processes. 

\begin{figure}[H]
	\centering
   \includegraphics[width=\textwidth]{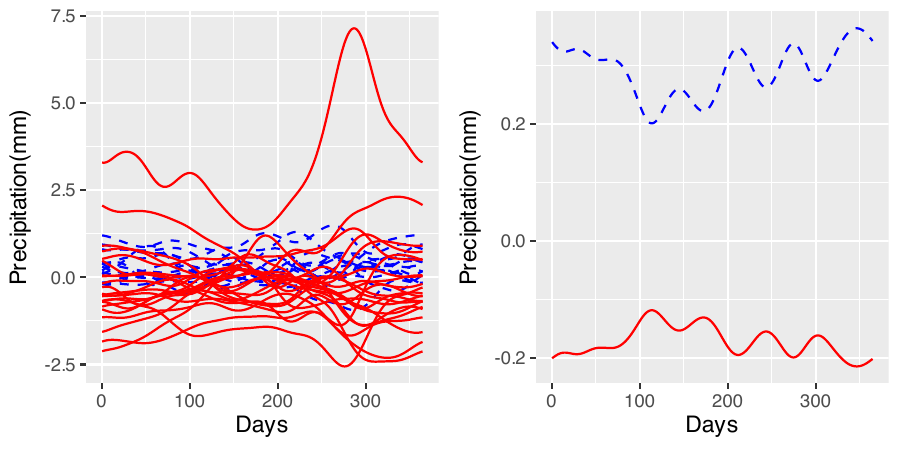}
   \caption{Left: profiles of residuals from linear function-on-function regression after and before the estimated structural break obtained from FF-dist. Right: the mean functions of the residuals after and before the estimated structural break. In both panels, the red solid and blue dashed lines represent the profiles from after and before the structural break, respectively.}
	\label{fig:Canada_residual}
\end{figure}

We employ the FPCA-based method proposed by \cite{yao2005b} to fit this model. This method can be implemented through the function FLM in the R package {\it fdapace} \citep{fdapace}. The number of FPCs that are used to estimate $\beta$ is selected through AIC. Then we apply the proposed procedure to $\hat{e} \lo i (t)$'s to test the null hypothesis. We obtained a $p$-value of 0.023 based on the null distribution of $T \lo n$ defined in \eqref{eq-KS}. 
This indicates strong evidence against this null hypothesis. Moreover, Figure \ref{fig:Canada_residual} displays the profiles of $\hat{e} \lo i$'s before and after the estimated structural break, and the corresponding mean functions. The stark difference in the two mean functions, as shown in the right panel of Figure \ref{fig:Canada_residual}, further justifies that model \eqref{eq:ffreg} does not fit the data reasonably well.

\section{Conclusion} \label{sec:conclusion}

In this article, we develop a fully functional method to detect and date structural breaks in the underlying distribution for independently observed functional data. This method is built upon mean embedding of functional data and the structure of nested Hilbert spaces, and is easy to implement in practice. 
Compared with existing work on structural breaks for functional data, 
the assumptions for our method are considerably weaker. In particular, our method does not rely on the classical functional change-point model \eqref{eq-DGP}, which is usually required  by existing methods. Moreover, we establish appealing theoretical properties of the proposed method under mild conditions. These theoretical guarantees greatly broaden the applicability of the proposed method. 
Through extensive simulation studies, 
we demonstrated that our method performs notably better than the existing methods when the structural break lies  in the aspects of  distribution of the functional data beyond the mean and the covariance functions. Furthermore, even in the cases where the structural break can be fully characterized by the mean and covariance functions, our method is not inferior to the existing  methods that are designed for such cases. Additionally, simulation studies and real examples show that our bootstrap confidence interval is much shorter than the conservative one developed in \cite{aue2018}, but still maintains a coverage probability close to the nominal level.

\bibliographystyle{apalike}
\bibliography{sfqr}
%\newpage 

\begin{comment}
\red{\begin{theorem}
    If $E \ka (X,X) < \infty$, then $\Sigma \lo {XX}$ is trace class.
\end{theorem}}

\red{\proof Since 
\begin{align*}
    \ka (\cdot, X) - \mu \lo X = \sum \lo {i=1} \hi \infty \langle \ka (\cdot, X) - \mu \lo X, \phi \lo i \rangle \lo {\frak M \lo X } \phi \lo i, 
\end{align*}
$\{ \phi \lo i \}$ is an orthonormal set, we have 
\begin{align*}
   \| \ka (\cdot, X) - \mu \lo X \| \hi 2 =\ali  \sum \lo {i=1 } \hi n \langle \ka (\cdot, X) - \mu \lo X, \phi \lo i \rangle \lo {\frak M \lo X } \hi 2 \\
=\ali  \sum \lo {i=1 } \hi n \langle \phi \lo i, \{ [\ka (\cdot, X) - \mu \lo X] \otimes [\ka (\cdot, X) - \mu \lo X] \} \phi \lo i   \rangle \lo {\frak M \lo X }.
\end{align*}
Since the  left  side can be rewritten as 
$
    \ka (X, X) - 2 \langle \ka (\cdot, X), \mu \lo X \rangle \lo {\frak M \lo X} - \| \mu \lo X \| \hi 2  
$, we have 
\begin{align*}
 \ka (X, X) - 2 \langle \ka (\cdot, X) , \mu \lo X \rangle \lo {\frak M \lo X}- \| \mu \lo X \| \hi 2 = \sum \lo {i=1 } \hi n \langle \phi \lo i, \{ [\ka (\cdot, X) - \mu \lo X] \otimes [\ka (\cdot, X) - \mu \lo X] \} \phi \lo i   \rangle \lo {\frak M \lo X }.
\end{align*}
Taking expectation on both sides, we have 
\begin{align*}
   E  \ka (X,X)  - \| \mu \lo X \| \lo {\frak M \lo X} \hi 2 
=  \sum \lo {i=1 } \hi n \langle \phi \lo i, \Sigma \lo {XX} \phi \lo i   \rangle \lo {\frak M \lo X } = \sum \lo {i=1} \hi \infty \lambda \lo i. 
\end{align*}
Since $E \ka (X,X) < \infty$,  the left-hand side is finite. Hence  the right-hand side is finite. \eop }
\end{comment}
\end{document}